\documentclass[twocolumn, 10pt]{extarticle} %'ext' is so we can get larger base font sizes

\usepackage[english]{babel}
\usepackage[utf8]{inputenc}
\usepackage[T1]{fontenc}
\newsavebox{\foobox}
\newcommand{\slantbox}[2][0]{\mbox{%
        \sbox{\foobox}{#2}%
        \hskip\wd\foobox
        \pdfsave
        \pdfsetmatrix{1 0 #1 1}%
        \llap{\usebox{\foobox}}%
        \pdfrestore
}}
\newcommand\unslant[2][-.25]{\slantbox[#1]{$#2$}}

\newcommand{\micron}{$\unslant\mu$m}

\usepackage[margin=0.7in]{geometry}
\usepackage{amsmath}
\usepackage{amssymb}
\usepackage{graphicx} % use png / jpg images
\usepackage{float}    % use [htb] float location
\usepackage[colorlinks=true, allcolors=blue]{hyperref}
\usepackage{authblk} %for author affiliations
\usepackage{wrapfig} % to let text flow around figure in abstract
\usepackage{cite}

\title{Multi-resonant high-$Q$ plasmonic metasurfaces}
\author[1,*]{Orad Reshef}
\author[2]{Md Saad-Bin-Alam}
\author[3]{Mikko J. Huttunen}
\author[4]{Graham Carlow}
\author[4]{Brian T. Sullivan}
\author[1]{Jean-Michel M\'{e}nard}
\author[1,2]{Ksenia Dolgaleva}
\author[1,2,5]{Robert W. Boyd}
\affil[1]{Department of Physics, University of Ottawa, 25 Templeton Street, Ottawa, ON K1N 6N5, Canada}
\affil[2]{School of Electrical Engineering and Computer Science, University of Ottawa, Ottawa, ON K1N 6N5, Canada}
\affil[3]{Photonics Laboratory, Physics Unit, Tampere University, P.O. Box 692, FI-33014 Tampere, Finland}
\affil[4]{Iridian Spectral Technologies Inc., 2700 Swansea Crescent, Ottawa, ON K1G 6R8, Canada}
\affil[5]{Institute of Optics and Department of Physics and Astronomy, University of Rochester, Rochester, NY 14627, USA}
\affil[*]{Corresponding author: orad@reshef.ca}

\date{\vspace{-3em}}

\begin{document}

\twocolumn[
  \begin{@twocolumnfalse}
    \maketitle
    
%%%%%%%%%%%%%%%%%%%%%%% ABSTRACT %%%%%%%%%%%%%%%%%%%%%%%
%%%%%%%%%%%%%%%%%%%%%%%%%%%%%%%%%%%%%%%%%%%%%%%%%%%%%%%%
    \begin{abstract}
      \vspace{0.5em}
      
      \hspace{-1.25cm}\begin{minipage}{0.4\textwidth}
      {\includegraphics[width=2.75in]{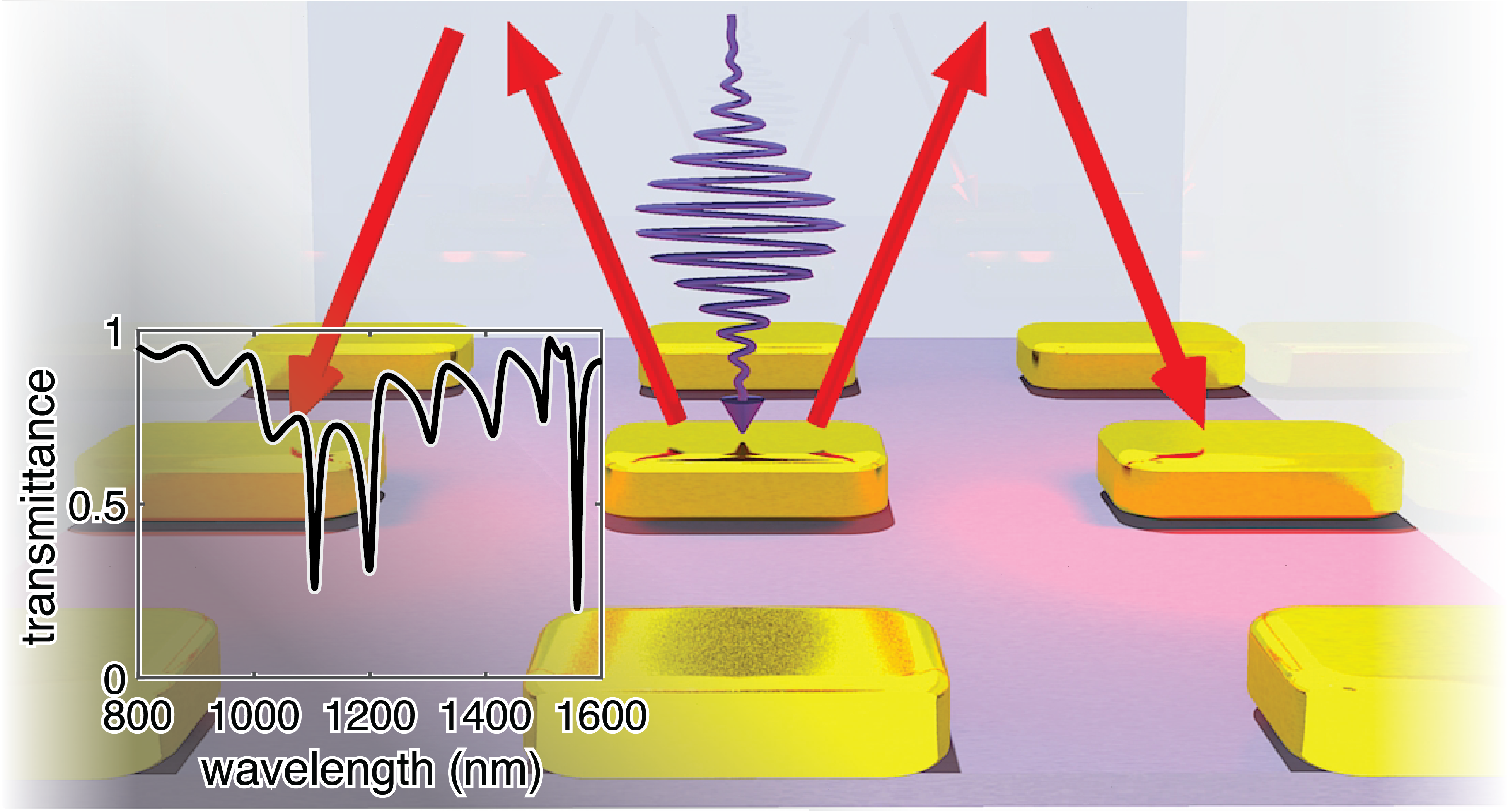}}
      \end{minipage}
      \begin{minipage}{0.6\textwidth}
    \noindent Resonant metasurfaces are devices composed of nanostructured sub-wavelength scatterers that generate narrow optical resonances, enabling applications in filtering, nonlinear optics, and molecular fingerprinting.  It is highly desirable for these applications to incorporate such devices with multiple, high-quality-factor resonances; however, it can be challenging to obtain more than a pair of narrow resonances in a single plasmonic surface. Here, we demonstrate a multi-resonant metasurface that operates by extending the functionality of surface lattice resonances, which are the collective responses of arrays of metallic nanoparticles. This device features a series of resonances with high quality factors ($Q \sim 40$), an order of magnitude larger than what is typically achievable with plasmonic nanoparticles, as well as a narrow free spectral range. This design methodology can be used to better tailor the transmission spectrum of resonant metasurfaces and represents an important step towards the miniaturization of optical devices. 
      \end{minipage}
    \end{abstract}
    
    \noindent\hfil\hspace{1cm}\rule{0.8\textwidth}{.4pt}\hfil
    \vspace{1em}
  \end{@twocolumnfalse}
]

%%%%%%%%%%%%%%%%%%%%%%%%% BODY %%%%%%%%%%%%%%%%%%%%%%%%%
%%%%%%%%%%%%%%%%%%%%%%%%%%%%%%%%%%%%%%%%%%%%%%%%%%%%%%%%

%\section{Introduction}
\noindent Plasmonic nanoparticles intrinsically support localized surface plasmon resonances (LSPRs) that may be spectrally located from the visible regime to the mid-infrared~\cite{Oldenburg1998, Maier2007}. These resonances can easily be tailored by modifying the nanoparticle geometry, and this design flexibility has opened up the field of nanoplasmonics for applications in filtering~\cite{Yokogawa2012, Zeng2013}, colour generation~\cite{Hsu2014,Kumar2012, Kristensen2016, Guay2017}, and, due to the large intrinsic nonlinearities of metals, nonlinear optics~\cite{Palomba2009, Li2017d, Alam2018}. More recently, gold nanoparticles have been used as meta-atom building blocks in metamaterials and metasurfaces that exploit the tunability of LSPRs~\cite{Schurig2006, Yu2011, Sun2012, Meinzer2014, Karimi2014}.

\begin{figure}[!t]
\begin{centering}
\includegraphics[width=1.05\linewidth]{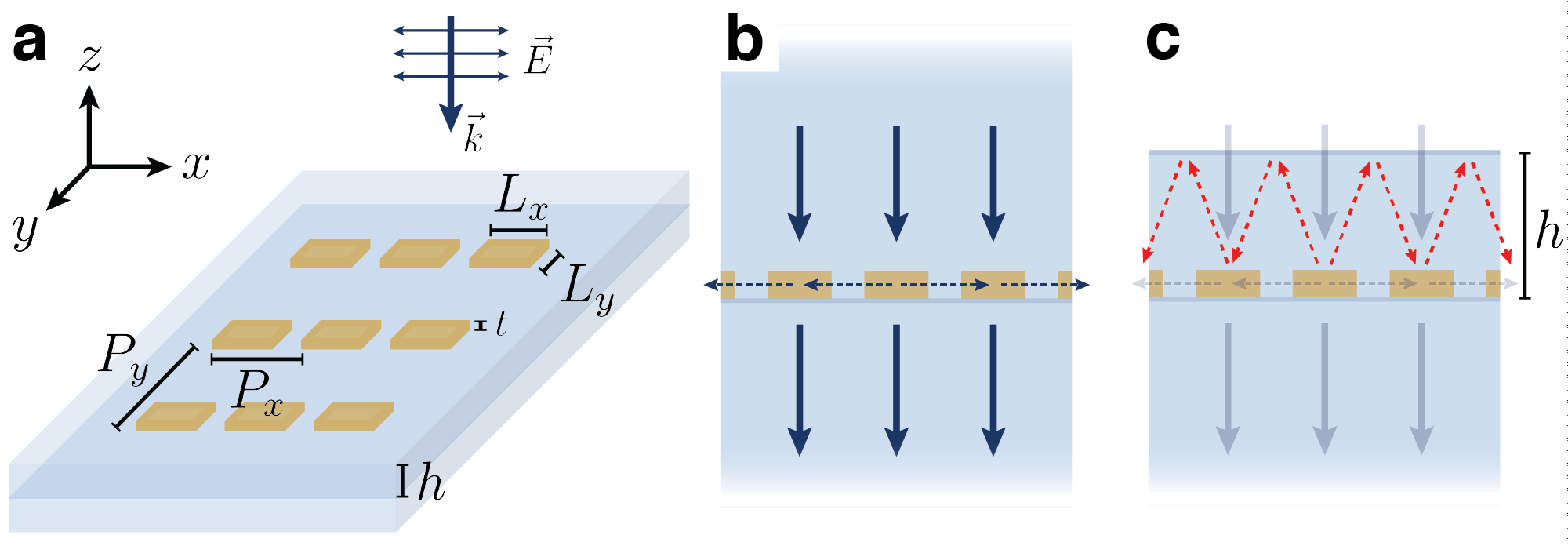}
\par\end{centering}
\caption{{\bf a)}~Schematic of the metasurface under investigation, consisting of rectangular gold nanoparticles in a rectangular array and cladded by a thin silica cladding layer.  {\bf b)} In a homogeneous medium, an incident plane wave (solid arrows) excites a surface lattice resonance. Here, every dipole is excited, scatters, and excites its neighbours  (dashed blue arrows). {\bf c)}~A thin cladding layer ($h$ on the order of a few wavelengths) confines additional resonance modes, providing another coupling channel between the dipoles (dashed red arrows). For clarity, the illustrated dashed arrows in parts b-c indicate the light scattered from only the middle incident arrow.}
\label{Fig:schematic}
\end{figure}

\begin{figure*}[!t]
\begin{centering}
\includegraphics[width=1\linewidth]{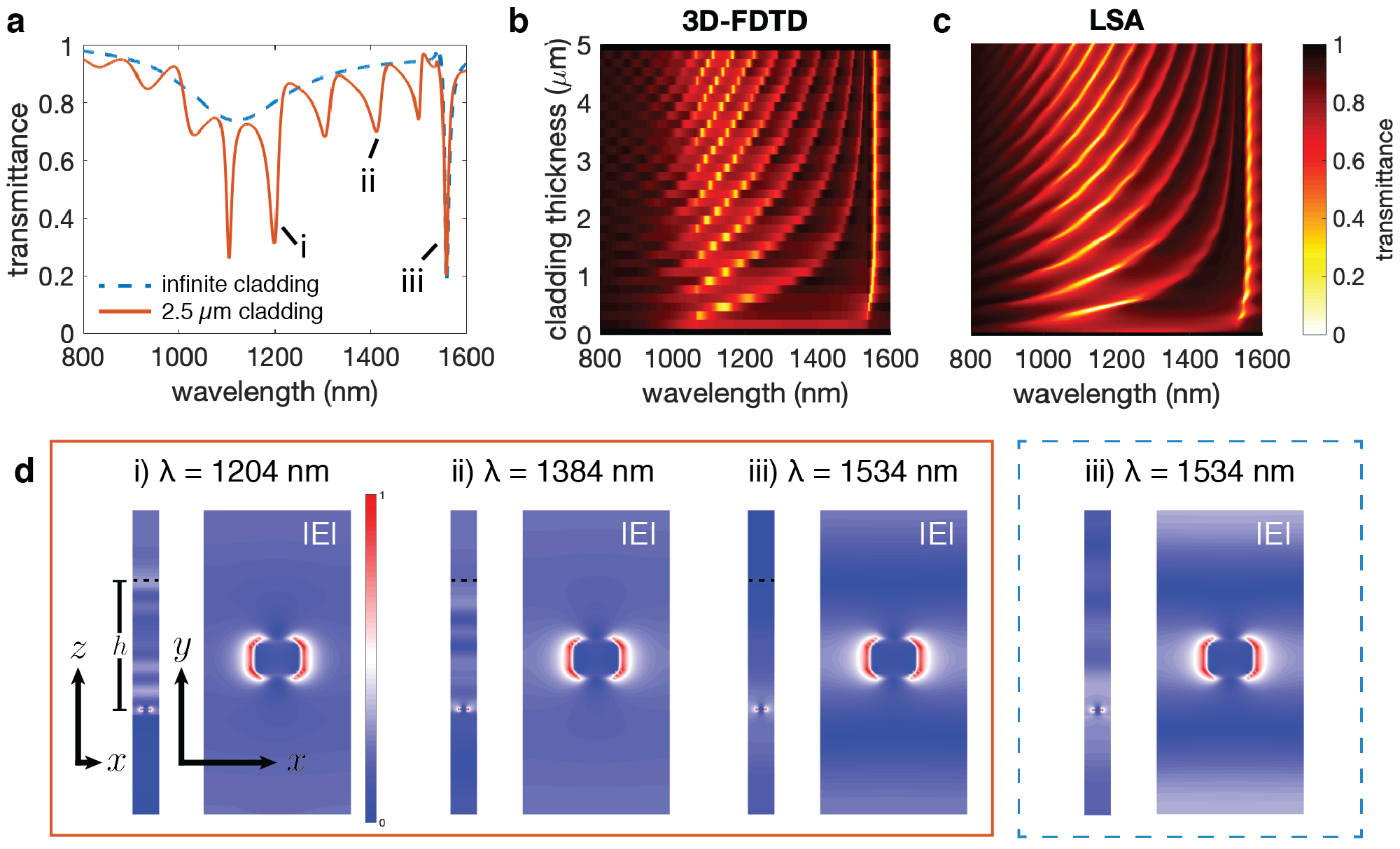}
\par\end{centering}
\caption{{\bf a)}~Full-wave simulations of transmission spectra of a periodic lattice of gold nanoparticles in a homogeneous environment (dashed line) and with a cladding layer of finite thickness (solid line). {\bf b)}~Transmission spectra as a function of cladding thickness calculated using 3D-FDTD. Additional resonances with higher quality factors appear as the thickness of the cladding increases. {\bf c)}~Corresponding transmission spectra calculated using the lattice sum approach. {\bf d)}~Near-field profiles of resonances indicated in (a) for a 2.5~\micron-thick cladding layer (solid red box) and for an infinitely thick cladding layer (dashed blue box).}
\label{Fig:simulations}
\end{figure*}

Despite this versatility, applications that feature plasmonic particles are limited by the inherent absorption loss of metals, which results in low quality factors ($Q\lesssim 10$) in the optical regime~\cite{Doiron2019, Kravets2018}. Recently, surface lattice resonances (SLRs) have emerged as an alternative method to obtain narrow-band resonances in plasmonic systems~\cite{Zou2004, Auguie2008, Chu2008, Kravets2018}. These resonances appear in surfaces consisting of periodic arrays of metal nanoparticles, and they feature $Q$ factors that are typically much higher than their LSPR counterparts, experimentally achieving quality factors on the order of $Q>100$~\cite{Li2014b, Yang2015a}, with numerical predictions exceeding 2000~\cite{Zakomirnyi2017}. These large quality factors are advantageous to many practical applications of plasmonic devices, particularly in nonlinear optics. Indeed, SLR metasurfaces have already been shown to enhance nonlinearities~\cite{Zaplicki2016, Michaeli2017, Huttunen2018, Huttunen2019a}, and they have been implemented in lasers~\cite{Zhou2013, Yang2015a, Hakala2017} and nonlinear spectroscopy applications~\cite{Hooper2019}.

Many applications have been shown to benefit strongly from the existence of multiple simultaneous resonances~\cite{Ali2019}, such as molecular fingerprinting~\cite{Choi2018}, fluorescence imaging~\cite{Liu2015}, heads-up display technologies~\cite{Hsu2014, SoljacicFullColorDisplay, Monti2017} or nonlinear applications, such as in frequency mixing~\cite{Thyagarajan2012, Aouani2012, Liu2016a} or frequency comb generation~\cite{Okawachi2011, Kippenberg2011}. Multiple simultaneous resonances are usually enabled by the hybridization of LSPRs such as in nanoparticle dimers and trimers~\cite{Nordlander2004, Thyagarajan2012}, by using asymmetric particles~\cite{Liu2015, Monti2017}, or through the use of multiple different materials~\cite{SoljacicFullColorDisplay}. More recent developments include the hybridization of multiple surface lattice resonances by incorporating multiple particles within a single unit cell~\cite{Baur2018}, among many other approaches in coupling both localized and delocalized responses~\cite{Ali2019}. However, to date high-$Q$ SLR metasurface designs have only supported one or two simultaneous SLRs~\cite{Baur2018, Guo2017a}, typically owing to the number of degrees of freedom in a two-dimensional plane~\cite{Guo2017a}. Here, we show that a metasurface consisting of a metallic nanoparticle array cladded by a thin transparent layer may exhibit multiple hybrid high-$Q$ SLR cavity modes.

%%%%%%%%%%%%%%%%%%%%%%%% THEORY %%%%%%%%%%%%%%%%%%%%%%%%
%%%%%%%%%%%%%%%%%%%%%%%%%%%%%%%%%%%%%%%%%%%%%%%%%%%%%%%%

%\section{Theory}

In a periodically arranged array of plasmonic meta-atoms (Fig~\ref{Fig:schematic}a), individual LSPRs strongly couple to form a hybridized collective lattice mode. This mode corresponds to the first-order diffraction mode of the array, which consists of a pair of counter-propagating surface modes in the plane of the array (Fig~\ref{Fig:schematic}b). The lattice response is often modeled using the coupled-dipole method (CDM) or the lattice-sum approach (LSA), which calculate the effective polarizability induced in a single particle of the array by taking into account the change in the local field felt by the particle due to the light re-scattered from all the other dipoles in the lattice~\cite{Zou2004, Auguie2008}. The explicit form of this lattice sum therefore depends strongly on the specific arrangement of the dipoles and their respective coupling channels (see Methods). For light at normal incidence to the surface and polarized along the $x$ axis, an orthogonal lattice periodicity $P_y$, and a homogeneous background refractive index of $n$, the LSA predicts that the resonance frequency of this mode is to be found at about $\lambda\approx n P_y $. The $Q$ factor and extinction ratio of this resonance depend on the polarizability (and, therefore, the dimensions) of the individual nanoparticles~\cite{Zakomirnyi2017}. The SLR wavelength has also been shown to depend on the nanoparticle dimension~\cite{Auguie2008}.

An interesting situation arises when a wavelength-scale transparent film is deposited on top of the periodic lattice. Figure~\ref{Fig:simulations}a compares the transmission spectra calculated using full-wave simulations for a metasurface composed of a rectangular array of gold nanoparticles in a homogeneous medium and a metasurface with a finite 2.5-\micron-thick silica (SiO$_2$) cladding. The particle dimensions are $L_x=200\mathrm{~nm}$ by $L_y=130\mathrm{~nm}$, and the thickness of the gold layer is only $t=20$~nm. The rectangular lattice parameters are $P_y=(1550\mathrm{~nm})/n\approx1060$~nm and $P_x=500$~nm. In the usual situation with an infinitely thick cladding layer, we expect only the broad ($Q\approx4$) LSPR with a resonance wavelength at 1100~nm and the narrow ($Q\approx 150$) SLR at $\lambda=1550$~nm to appear. However, when there is a finite cladding thickness, a plurality of hybrid resonance modes emerge. In particular, the SLR remains unchanged, and the hybrid resonances nearest to the nanoparticle's LSPR wavelength have both the highest $Q$ factors and extinction ratios. The $Q$ factor (as estimated from the full-width at half maximum) of the SLR remains at about 150, and the $Q$ factor for the narrowest hybrid resonance is about 100.

To further investigate the origin of these new resonances, we perform additional numerical calculations (Figs.~\ref{Fig:simulations}b~--~c). First, we perform finite-difference time-domain simulations for various top-cladding thicknesses $h$. Without any cladding (\emph{i.e.,} $h=0$), the SLR mode is not supported, as expected~\cite{Auguie2008, Kravets2018}. As we gradually increase $h$, we also observe an increasing number of resonances in the transmission spectrum. Additionally, the most pronounced resonances continue to appear near the nanoparticle resonance, around $\lambda=1100$~nm. Next, we perform the same calculations using the LSA. Here, the effect of a finite cladding layer can be taken into account by adding a coupling term between the dipoles due to reflection occurring at the superstrate-air interface (see Fig~\ref{Fig:schematic}c). The LSA model presented in Fig.~\ref{Fig:simulations}c is in excellent agreement with the full-wave simulations, confirming that the origin of these additional modes may be attributed solely due to this coupling channel. Therefore, in addition to surface waves that travel parallel to the plane of the metasurface, the finite cladding layer supports modes that reflect diagonally between the metasurface and the upper cladding interface.

Figure~\ref{Fig:simulations}d illustrates cross-sectional profiles of the magnitude of the electric field $|E|$ for a select few resonances. In all cases, the field is most intense on the surface of the nanoparticle. When looking out of the plane, we observe a notable amount of field also bound by the cladding layer away from the particle. Figure~\ref{Fig:simulations}d also shows the field profile for the original SLR in a homogeneous background. It is found to be nearly identical to its corresponding resonance with a finite cladding layer. This behaviour is expected since in both cases, this resonant mode is propagating parallel to the plane of the surface, and it helps explain why its resonance wavelength does not change substantially as a function of cladding thickness.

%%%%%%%%%%%%%%%%%%%%%% EXPERIMENT %%%%%%%%%%%%%%%%%%%%%%
%%%%%%%%%%%%%%%%%%%%%%%%%%%%%%%%%%%%%%%%%%%%%%%%%%%%%%%%
%\section{Experiment}
\begin{figure}[!b]
\begin{centering}
\includegraphics[width=1\linewidth]{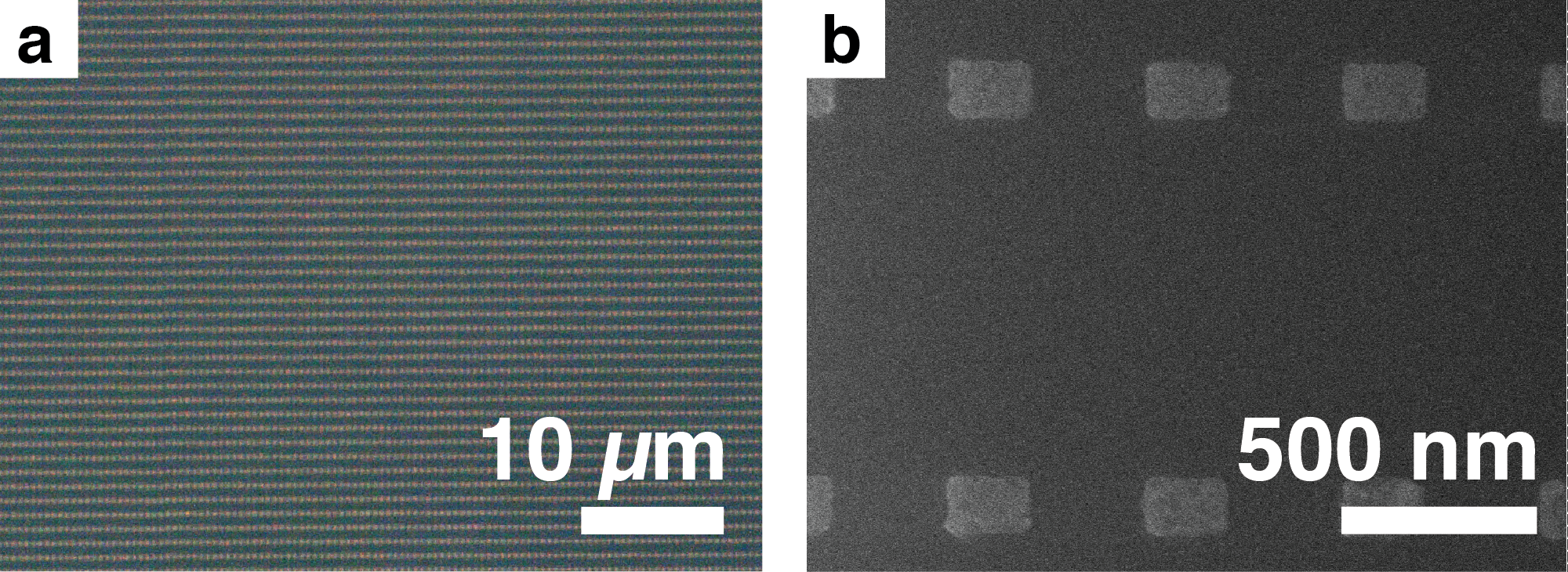}
\par\end{centering}
\caption{ {\bf a)}~Bright-field optical microscope image of the metasurface. {\bf b)}~Helium ion microscope image of the metasurface prior to cladding deposition.}
\label{Fig:fab}
\end{figure}

\begin{figure*}[!htb]
\begin{centering}
\includegraphics[width=1\linewidth]{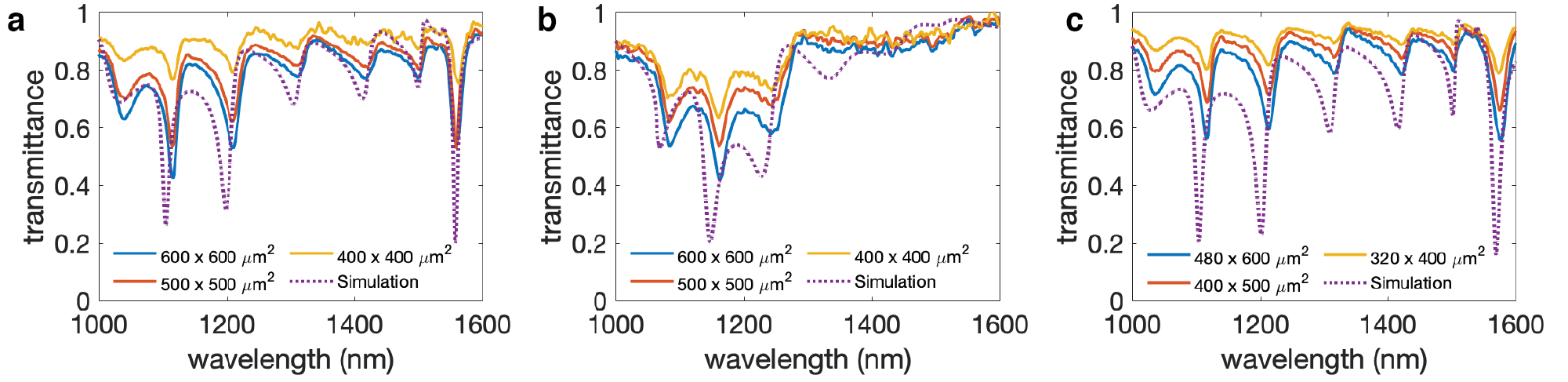}
\par\end{centering}
\caption{Transmission spectra for {\bf (a)}~$P_x=500$~nm by $P_y=1.06$~\micron{}, {\bf (b)}~$P_x=1.06$~\micron{} by $P_y=500$~nm, and {\bf (c)}~$P_x=400$~nm by $P_y=1.06$~\micron{}. The dashed line corresponds to full-wave simulations with periodic boundary conditions. Multiple resonances emerge with quality factors that are significantly larger than those of the individual nanoparticles. No SLR is supported at $\lambda=1550$~nm when $P_y\neq 1.06$~\micron, as expected. The measurements are taken for arrays of different sizes --- the extinction is observed to increase monotonically for larger arrays.}
\label{Fig:measurements}
\end{figure*}

To experimentally confirm these new hybrid resonances, we fabricate a series of arrays of this proposed device (\emph{i.e.,} $P_x=500$~nm by $P_y=1060$~nm) with a 2.5-\micron-thick silica top-cladding and with lattice areas of $400\times400$~\micron$^2$, $500\times 500$~\micron$^2$, and $600\times600$~\micron$^2$.  Figure~\ref{Fig:fab} shows images of the metasurface before the final cladding deposition step. To isolate the effect of the lattice configuration from that of the nanoparticle dimensions, we also fabricated a matching set of devices with identical nanoparticles in the same orientation, but in a rotated lattice configuration (\emph{i.e.,} $P_x=1060$~nm by $P_y=500$~nm). Finally, we also fabricated another different series of devices that support SLR at $\lambda=1550$~nm (\emph{i.e.,} $P_x=400$~nm by $P_y=1060$~nm). In these devices, the particle number was held constant, resulting in total array sizes of $320\times400$~\micron$^2$, $400\times500$~\micron$^2$, and $480\times600$~\micron$^2$.

Figure~\ref{Fig:measurements} displays the transmittance through these devices as a function of wavelength. The resonances are present in all devices, in good agreement with the simulations. In devices where $P_y=1060$~nm, the SLR appears around the design wavelength of $\lambda=1550$~nm. The resonances with the highest extinction ratio appear near the LSPR, as predicted by the simulations. Notably, the sharpest hybrid resonance has a $Q$ factor of 40, an order of magnitude larger than that of the individual nanoparticles. In every device, both the extinction ratios and the $Q$ factors of the resonances increase as a function of array size. Finally, in the array with the rotated lattice configuration, the SLR does not appear at $\lambda=1550$~nm, as expected (Fig.~\ref{Fig:measurements}b).

%\section{Discussion and conclusions}
The plasmonic metasurface described above supports multiple resonances with $Q$ factors that are an order of magnitude larger than those associated with the LSPRs of the individual nanoparticles. The resulting hybrid modes propagate along trajectories described by the LSA method, and may be interpreted as higher-order diffraction modes that are confined by the upper boundary~\cite{Wang1993}. 
Full-wave simulations show that, despite their differing trajectories, the different resonances possess similar field profiles on the surface of the nanoparticle, with good spatial overlap. These types of multi-resonant surfaces could be exploited for applications in optical frequency mixing, which would benefit from the combination of the strong nonlinearities of metals, strong field overlap, and the modification of the local density of states of high quality resonances. The location and number of resonances is determined by the lattice spacing and the thickness of the top-cladding layer, and can be rapidly predicted using the modified LSA. As the array size increases, the devices perform better, manifesting higher $Q$ factors and deeper extinction ratios. We note that the simulation, which treats an ``infinitely periodic'' array, exhibits the narrowest and deepest resonances, suggesting that larger arrays could display even better performance, as has been explored in Ref.~\cite{Zundel2018}.

The effect of combining a Fabry-P\'{e}rot (FP) microcavity with plasmonic nanoparticles has already been thoroughly investigated in the literature~\cite{Schmidt2012a, Yao2014, Li2013b, Ramezani2016}. Typically, lattice parameters are deeply sub-wavelength such that the metasurface may be treated as an effective reflective boundary in an asymmetric FP cavity. This modeling approach does not work in our case due to the emergence of diffraction orders. In a more recent implementation~\cite{Ramezani2016}, an SLR is hybridized with the propagating eigenmodes of a higher-index slab waveguide. In our approach, the cladding layer possesses the same refractive index as the substrate, and so would not support any propagating slab modes in the absence of the metasurface. Therefore, our method, which capitalizes on the periodic structure of the lattice to exploit the highest-$Q$ resonances, is based on a fundamentally novel resonance mechanism. Its functionality can also be further expanded by combining with other multi-resonance mechanisms, such as by employing unit cells that contain more than a single nanoparticle~\cite{Baur2018}.

One interesting consequence of this resonance mechanism is that the free spectral range (FSR) of this device may turn out to be much narrower than that of an equivalent FP resonator. An FP etalon formed of 2.5~\micron{} of silica glass is predicted to have an FSR of $\Delta \nu_{\mathrm{FSR}}=\frac{c}{2n_gL}=42\mathrm{~THz}$ (corresponding to $\Delta \lambda_{\mathrm{FSR}}=185\mathrm{~nm}$ near $\lambda=1150\mathrm{~nm}$). However, our devices all demonstrate a much narrower FSR of $\Delta \nu_{\mathrm{FSR}}=22.8\mathrm{~THz}$ (\emph{i.e.,} $\Delta \lambda_{\mathrm{FSR}}=94\mathrm{~nm}$). This platform therefore provides a method with which one may relax the restrictions on device size for a given desired free spectral range.

If what is actually desired is to suppress this multi-resonance mechanism, our work tells us that the cladding layer needs to be much thicker than what is typically used, for example, to enable high-quality photonic integrated circuits (\emph{e.g.,} Ref.~\cite{Ji2017}). Indeed, the simulations in Fig.~\ref{Fig:simulations} show that a wavelength-scale cladding layer is insufficiently thick to suppress the additional modes, and that the presence of this layer needs to be taken into account for SLR metasurfaces with a finite cladding.

The approach presented above can be used to design the highest $Q$-factor  multi-resonant metasurfaces that incorporate plasmonic materials, which is desirable for nonlinear applications. These devices feature a simple fabrication procedure, can be composed using common, non-exotic materials, and may be trivially scaled to other operating wavelengths of interest, particularly at longer wavelengths where dielectric stacks become prohibitively thick for standard deposition methods. 

%%%%%%%%%%%%%%%%%%%%%%% METHODS %%%%%%%%%%%%%%%%%%%%%%%%
%%%%%%%%%%%%%%%%%%%%%%%%%%%%%%%%%%%%%%%%%%%%%%%%%%%%%%%%
\section*{Methods}

\subsection*{Simulations}
Full-wave simulations were performed using a commercial three-dimensional finite-difference time-domain (3D-FDTD) solver. A single unit cell was simulated using periodic boundary conditions in the in-plane dimensions and perfectly matched layers in the out-of-plane dimension. The structures were created using fully dispersive optical material properties for silica~\cite{Palik} and for gold~\cite{Johnson1972}. Minimal artificial absorption ($\mathrm{Im}(n)\,{\sim}\,10^{-4}$) was added to the background medium to reduce numerical divergences.

The lattice sum approach is a simplified variant of the discrete-dipole approximation method~\cite{Draine1994}. The version of the LSA code that was used to compute the figures in this manuscript can be found at Ref.~\cite{LSAcode}. The LSA differs from the DDA in that it assumes an infinite array of identical dipoles ($\vec{p}_i=\vec{p}_k=\vec{p}$)~\cite{Huttunen2016a}. Under these assumptions, the dipole moment $\vec{p}$ of any particle in the array can be written as  
\begin{equation} \label{Eq:p_lsa}
\vec{p} = \frac{\epsilon_0 \alpha \vec{E}_\mathrm{inc}}{1 - \alpha \mathcal{S}} \equiv \epsilon_0 \alpha^* \vec{E}_\mathrm{inc}, 
\end{equation}
where the effect of inter-particle coupling is incorporated in the lattice sum $\mathcal{S}$, and $\alpha^*$ is the effective polarizability. For arrays with in-plane coupling, the lattice sum term $\mathcal{S}_{\mathrm{in}}$ is simply
\begin{align} \label{Eq:S_in}
\mathcal{S}_{\mathrm{in}} =& \sum\limits^N_{j=1} \frac{\mathrm{exp}(\mathrm{i}kr_j)}{r_{j}} \Bigg[ k^2 \sin^2\theta_{j}  \nonumber\\
& +\frac{(1-\mathrm{i}kr_j )(3 \cos^2\theta_{j}-1)}{r_j^2} \Bigg],  
\end{align}
where the sum extends over the $N$ nearest neighboring dipoles, $r_j$ is the distance to the $j^{\mathrm{th}}$ dipole, and $\theta_j$ is the angle between $\mathbf{r}_j$ and the dipole moment $\vec{p}$. The simulation in Fig.~\ref{Fig:simulations} considers all $N=680,000$ particles found in the largest (\emph{i.e., } $600\times600$~\micron$^2$) array.

For the situation where particles may also optically couple back into the array via scattered fields reflected at the superstrate-air interface, the total lattice sum term is modified by an additional contribution:
\begin{equation}
    \mathcal{S}=\mathcal{S}_{\mathrm{in}}+\mathcal{S}_{\mathrm{FP}}.
\end{equation}
The lattice sum term $\mathcal{S}_{\mathrm{FP}}$ for arrays with out-of-plane (Fabry-P\'{e}rot type) coupling takes the form 
\begin{align} \label{Eq:S_FP}
\mathcal{S}_{\mathrm{FP}} =& \sum\limits^N_{j=1} \frac{R_j \mathrm{exp}(\mathrm{i}k d_j)}{d_j} \Bigg[ k^2 \sin^2\theta_{j}  \nonumber\\
& +\frac{(1-\mathrm{i}k d_j)(3 \cos^2\theta_{j}-1)}{d^2_j} \Bigg],
\end{align}
where $h$ is the thickness of the upper cladding, $d_j = 2\sqrt{r_j^2/4+h^2}$, and $R_j$ are the appropriate Fresnel amplitude reflection coefficients. A similar approach to this structure can be found in Refs.~\cite{Schokker2017,Chen2017a}.
Once the particle polarizabilities are known, the optical extinction spectra can be obtained by solving for Eqs.~\eqref{Eq:S_in}--\eqref{Eq:S_FP} and using the optical theorem, $ \mathrm{Ext} \propto k \mathrm{Im}(\alpha^*)$~\cite{Jackson}.

\subsection*{Fabrication}
The metasurfaces are fabricated using a standard metal lift-off process and a positive tone resist bi-layer. We start with a fused silica substrate, and define the pattern using electron-beam lithography with the help of a commercial conductive polymer. The mask was designed using shape-correction proximity error correction~\cite{Schulz2015a} to correct for corner rounding. Following development, gold is deposited using thermal evaporation. The final silica cladding layer is deposited using sputtering. The backside of the silica substrate is coated with an anti-reflective coating to minimize substrate-related etalon fringes.

\subsection*{Characterization}
The samples are excited using a collimated tungsten-halogen light source. The incident polarization is controlled using a broadband linear polarizing filter. The entire sample is illuminated, and the transmission from a single device is selected using a variable aperture. The transmission spectrum is measured using a diffraction-grating-based optical spectrum analyzer, and is normalized to a background trace of the substrate without gold nanostructures.

\section*{Supporting Information Available}
The version of the code used to generate the figures in this manuscript is provided in the supporting information. Up-to-date LSA FP code is provided for free at doi:\href{https://doi.org/10.5281/zenodo.3259084}{10.5281/zenodo.3259084}. Users of this code are kindly requested to acknowledge and cite its original description and use in this work.

\section*{Author contributions}
OR and MSBA conceived the basic idea for this work. OR performed the FDTD simulations; MJH performed the lattice sum calculations. OR and GC fabricated the devices. MSBA built the experimental setup and carried out the measurements. OR analysed the experimental results. BS, JMM, KD, and RWB supervised the research and the development of the manuscript. OR wrote the first draft of the manuscript, and all authors subsequently took part in the revision process and approved the final copy of the manuscript. Portions of this work were presented at the Conference on Lasers and Electro-Optics in 2019~\cite{Saad-Bin-Alam}.

\section*{Acknowledgements}
Fabrication in this work was performed in part at the Centre for Research in Photonics at the University of Ottawa (CRPuO). The authors thank Anthony Olivieri for help with nanofabrication, M. Zahirul Alam for help with measurements, Choloong Hahn for help with imaging, and Mikhail A. Kats and Martti Kauranen for helpful discussions.

The authors acknowledge support from the Canada Excellence Research Chairs Program, the Canada Research Chairs Program, and the Natural Sciences and Engineering Research Council of Canada (NSERC) Discovery funding program. OR acknowledges the support of the Banting Postdoctoral Fellowship of the NSERC. MSBA acknowledges the support of the Ontario Graduate Scholarship and the University of Ottawa Excellence Scholarship. MJH acknowledges the support of the Academy of Finland (Grant No. 308596) and the Flagship of Photonics Research and Innovation (PREIN) funded by the Academy of Finland (Grant No. 320165).

\footnotesize
%\bibliography{library}

\end{document}